\documentclass[12pt]{article}

\def\ZZ {\bf Z}
\def\RR {\bf R}

\def\exp{{\rm exp}}
\def\etal{{\it et al.}}

\usepackage{fullpage,epsfig,fancyheadings}

\usepackage[psamsfonts]{euscript}

\usepackage{epic}

\usepackage{amstex}

\newcommand{\beq}{\begin{equation}}
\newcommand{\eeq}{\end{equation}}

\begin{document}
\vspace*{-.6in}
\thispagestyle{empty}

\begin{flushright}
CALT-68-2013\\
hep-th/9508143
\end{flushright}
\baselineskip = 20pt

\vspace{.5in}

{\Large
\begin{center}
An SL(2,Z) Multiplet of Type IIB Superstrings\footnote{Work 
supported in part by
the U.S. Dept. of Energy under Grant No. DE-FG03-92-ER40701.}
\end{center}}

\vspace{.4in}

\begin{center}
John H. Schwarz\\
\emph{California Institute of Technology, Pasadena, CA  91125, USA}
\end{center}
\vspace{1in}

\begin{center}

\textbf{Abstract}

\end{center}

\begin{quotation}
\noindent  An $SL(2, \ZZ)$ family of string solutions of type IIB supergravity
in ten dimensions is constructed.  The solutions are labeled by a pair of
relatively prime integers, which characterize charges of the three-form field
strengths.  The string tensions depend on these charges in an $SL(2,\ZZ)$
covariant way.  Compactifying on a circle and identifying with 
eleven-dimensional
supergravity compactified on a torus implies that the modulus of the IIB
theory should be equated to the modular parameter of the torus.
\end{quotation}
\vfil

\newpage
\pagenumbering{arabic}

Among the various conjectured duality symmetries of superstring theories, the
proposed $SL(2,\ZZ)$ symmetry of the type IIB superstring theory in ten
dimensions is especially interesting~\cite{hulla,greena}. Like the $SL(2,\ZZ)$
S duality of the $N=4$ 
$D=4$ heterotic string~\cite{font,sen}, it relates weak and
strong coupling.  However, unlike the heterotic example in which the symmetry
relates electrically and magnetically charged states of the same gauge field,
the IIB duality relates electrically charged states of two different gauge
fields.  In this respect it is more like a T duality \cite{giveon}.  Combined
with ordinary T dualities, the IIB $SL(2,\ZZ)$ duality implies the complete U
duality symmetry of toroidally compactified type II strings in dimensions
$D < 10$ \cite{hulla,witten}.

The $SL(2,\ZZ)$ duality of the IIB theory will be explored here by considering
string-like (or `one-brane') 
solutions of the $D=10$ IIB supergravity theory.  It will be
argued that there is an infinite family of such solutions 
forming an $SL(2,\ZZ)$
multiplet. (This possibility was hinted at in section 5 of Ref. \cite{harvey}.)
Once these string solutions have been constructed, we will consider
compactification on a circle and compare the resulting 
nine-dimensional spectrum with that of
eleven-dimensional supergravity compactified on a torus.  
The conclusion will be that the $SL(2,\ZZ)$ duality
group of the IIB theory corresponds precisely to the modular
group of the torus.

All ten-dimensional supergravity theories contain the following terms in common
\begin{equation}
{S}_0 = {1\over 2\kappa^2} 
\int d^{10} x \sqrt{-g} \left(R - {1\over 2}
(\partial \phi)^2 - {1\over 12} e^{-\phi} H^2\right),
\end{equation}
where $H$ is a three-form field strength $(H = dB)$, and $\phi$ is the dilaton.
Moreover, in each case, a solution to 
the classical equations of motion derived
from $S_0$ can be regarded as a solution 
of the complete supergravity theory with
all other fields set equal to zero.  A macroscopic string-like solution,
which was identified with the heterotic string,
was constructed by Dabholkar \etal\ \cite{dabholkara}  (This was
generalized to p-branes in Ref.~\cite{horowitz}.)  More recently, Dabholkar~\cite{dabholkarb} and
Hull~\cite{hullb} noted that, since the type I superstring has the same
low-energy effective action as the O(32) heterotic string, 
and since the zero modes work out suitably,
these macroscopic heterotic strings can also be
regarded as solutions of the type I theory.  This observation is consistent
with the proposal that the O(32) heterotic string at weak coupling is
equivalent to the type I superstring at strong coupling (and vice
versa)~\cite{witten}.  In this paper, the interest in
macroscopic string solutions will be in the context of the IIB theory.

The macroscopic 
string solution of Ref. ~\cite{dabholkara}, restricted to $D = 10$, is
given by
\renewcommand{\theequation}{2a}
\begin{equation}
ds^2 = A^{-3/4} [-dt^2 + (d {x}^1)^2] + A^{1/4} d {{\bf x}} \cdot
d{{\bf x}},
\end{equation}
\renewcommand{\theequation}{2b}
\begin{equation}
B_{01} = e^{2\phi} = A^{-1},
\end{equation}
where
\renewcommand{\theequation}{\arabic{equation}}
\setcounter{equation}{2}
\begin{equation}
A = 1 + {Q\over 3r^6} ,
\end{equation}
${\bf x} = ({ x}^2, {x}^3, ... , {x}^9)$, ${{\bf x}} \cdot
{{\bf x}} = r^2 = \delta_{ij} {x}^i {x}^j$, and $Q$ is the
$B_{\mu\nu}$ electric charge carried by the string.  (Recall that 
the electric charge of a $(p +
2)$-form field strength is carried by a p-brane.)  Strictly
speaking, the $S_0$ equations are not satisfied at $r = 0$, the string
location, because $\nabla^2 A$ has a delta-function singularity there.  In Ref.
{}~\cite{dabholkara} it was proposed that this could be fixed by 
coupling to a string source, which means considering $S
= S_0 + S_\sigma$ instead, where
\begin{equation}
S_\sigma = - {T\over 2} \int d^2 \sigma (\partial^\alpha X^\mu
\partial_\alpha X^\nu G_{\mu\nu} + \ldots),
\end{equation}
$T$ is the string tension, and
\begin{equation}
G_{\mu\nu} = e^{\phi/2} g_{\mu\nu}
\end{equation}
is the string metric.  From the coefficient of the delta function one can
deduce the relation $Q = {{\kappa}^2 T/ \omega_7}$, 
where $\omega_7 = {1\over 3}
\pi^4$ is the volume of $S^7$.\footnote{The metric (2a) has a naked 
singularity
at $r = 0$, though it corresponds to an extremal limit of a ``black string''
with the singularity behind a horizon~\cite{horowitz}.  Such borderline
singularities might be removed (or shielded) by corrections due to
higher-dimension terms in the effective field theory.  These corrections would
not change the relation between the tension and the charge, which follows from
saturation of a Bogomol'nyi bound.}

The type IIB theory has two three-form field strengths $H^{(i)} = dB^{(i)}$, 
$i = 1,2$~\cite{greenb}.  $H^{(1)}$ belongs to the NS-NS sector and can be
identified with $H$ in the preceding discussion.  $H^{(2)}$ belongs to the
R-R sector and does not couple to the (usual) string world sheet.  In
addition, the type IIB theory has two scalar fields, which can be combined into
a complex field $\lambda = \chi + ie^{-\phi}$.  The dilaton $\phi$ is in the
NS-NS sector and can be identified with $\phi$ in the preceding discussion,
while $\chi$ belongs to the R-R sector.  The other bose fields are the metric
$g_{\mu\nu}$ and a self-dual five-form field strength $F_5$.  The five-form
field strength (as well as all fermi fields) will be set to zero throughout
this paper.
The reason is that the corresponding charges are carried by a self-dual
three-brane, whereas the focus here is on charges carried by strings.  Once we
set $F_5 = 0$, it is possible to write down a covariant action that gives the
desired equations of motion~\cite{hullb}:
\begin{equation}
S = {1\over 2{\kappa}^2} 
\int d^{10} { x} \sqrt{-g} (R + {1\over 4} tr (\partial
\mathcal{M} \partial \mathcal{M}^{-1}) - {1\over 12} H^T \mathcal{M} H).
\end{equation}
Here we have combined $H^{(1)}$ and $H^{(2)}$ into a two-component vector $H =
dB$, and introduced the symmetric $SL(2,\RR)$ matrix
\begin{equation}
\mathcal{M} = e^\phi
\left( \begin{array}{cc} | \lambda |^2 & \chi\\
\chi & 1 \end{array} \right).
\end{equation}
This action has manifest invariance under the global $SL(2,\RR)$ transformation
\begin{equation}
\mathcal{M} \rightarrow \Lambda \mathcal{M} \Lambda^T, \quad B \rightarrow
(\Lambda^T)^{-1} B.
\end{equation}
The choice $\Lambda = \left(\begin{array}{cc} a & b\\ c & d \end{array}
\right)$ corresponds to

\renewcommand{\theequation}{9a}
\begin{equation}
\lambda \rightarrow {a \lambda + b\over c\lambda + d} \qquad
\end{equation}
\renewcommand{\theequation}{9b}
\begin{equation}
B^{(1)} \rightarrow d B^{(1)} - c B^{(2)}, \qquad
B^{(2)} \rightarrow a B^{(2)} - b B^{(1)}.
\end{equation}
\renewcommand{\theequation}{\arabic{equation}}
\setcounter{equation}{9}

Given the symmetry of this system, it is clearly artificial to only consider
solutions carrying $H^{(1)}$ electric charge and not $H^{(2)}$ electric charge.
Measured in units of $Q$, we will consider solutions carrying charges $(q_1,
q_2)$. Since there exist five-brane solutions carrying magnetic $H$ charge,
the generalized Dirac quantization condition~ \cite{nepomechie}
implies that $q_1$ and $q_2$ must be integers.  Moreover, $q_1$ and $q_2$
should be relatively prime,  since otherwise the solution is neutrally stable
against decomposing into a multiple string solution --- the number of strings
being given by the common divisor.\footnote{This is the same counting rule
that was required in a different context in Ref. \cite{strominger}.  
We will show that it leads to sensible degeneracies after compactification
on a circle. As was pointed out in Ref. \cite{strominger}, 
a different rule is sometimes appropriate in other situations.}
Also, the $(q_1, q_2)$ string and the
$(-q_1, -q_2)$ string are related by orientation reversal $({x}^1
\rightarrow -{x}^1)$.
A complete description of string solutions requires specifying
the vacuum in which they
reside.  In the IIB theory this means choosing 
the asymptotic value of $\lambda$ as
$r \rightarrow \infty$, denoted by $\lambda_0$.  The choice that is simplest,
and therefore will be considered first, is $\lambda_0 = i$, which corresponds
to $\chi_0 = \phi_0 = 0$.  The tension of the $(q_1, q_2)$ string,
denoted by $T_q$, will be determined. 

Replacing $T$ by $T_q$ in the solution described above still gives a
classical solution (for coupling a string of tension $T_q$ rather than $T$),
though in general it violates the quantization condition.  To describe
this solution, we replace $A(r)$ by $A_q (r)$, where
\begin{equation}
A_q(r) = 1 + {\alpha_q\over 3r^6}.
\end{equation}
Now consider letting $\alpha_q = \sqrt{q_1^2 + q_2^2}~ Q$ and $T_q =
\sqrt{q_1^2 + q_2^2}~ T$.  For this choice, we can recover the quantization
condition by applying an $SL(2,\RR)$ transformation given by
\begin{equation} \Lambda = {1\over\sqrt{q_1^2 + q_2^2}} 
\left( \begin{array}{cc}
q_1 & -q_2\\
q_2 & q_1 \end{array} \right)
\end{equation}
to obtain a new classical solution.  We have chosen $\Lambda$ to belong to the
$SO(2)$ subgroup of $SL(2,\RR)$, 
because this  preserves the modulus $\lambda_0 =
i$.  In this way we obtain a solution with charges $(q_1, q_2)$ given by
\begin{equation} \label{metric}
ds^2 = A_q^{-3/4} (-dt^2 + (d{ x}^1)^2) + A_q^{1/4} d{{\bf x}} \cdot d
{{\bf x}}\qquad ~~ \quad \qquad
\end{equation}
\begin{equation}
B_{01}^{(i)} = {q_i\over\sqrt{q_1^2 + q_2^2}} A_q^{-1}\qquad \qquad \qquad
\qquad \qquad \qquad~~ \qquad \qquad
\end{equation}
\begin{equation} \label{lambdaformula}
\lambda = {iq_1 A_q^{1/2} - q_2\over iq_2 A_q^{1/2} + q_1} = {q_1 q_2 (A_q -1)
+ i (q_1^2 + q_2^2) A_q^{1/2}\over q_1^2 + q_2^2 A_q} \ .
\end{equation}
It is evident that as $r \rightarrow \infty$, $A_q \rightarrow 1$, and
therefore $\lambda \rightarrow i$.

The next step is to generalize this family of solutions to an arbitrary vacuum
modulus $\lambda_0$.  For this purpose we start with arbitrary $\alpha_q$,
$\lambda_0 = i$, $B^{(1)} = \cos \theta A_q^{-1}$, 
$B^{(2)} = \sin \theta A_q^{-1}$,
and $\lambda = (i \cos \theta A_q^{1/2} - \sin \theta)/(i \sin \theta A_q^{1/2}
+ \cos \theta)$.  (The solution in eq.~(\ref{lambdaformula}) corresponds to the choice
$e^{i\theta} = (q_1 + iq_2)/\sqrt{q_1^2 + q_2^2}$.)  Next we apply the $SL(2,
\RR)$ transformation
\begin{equation}
\Lambda = \left( \begin{array}{cc}
e^{-\phi_{0}/2} & \chi_0 e^{\phi_{0}/2}\\
0 & e^{\phi_{0}/2} \end{array} \right),
\end{equation}
which maps $i \rightarrow \lambda_0$.  Finally, we must
satisfy the quantization condition.  This is achieved by the choice
\begin{equation}
e^{i\theta} = e^{\phi_0/2} (q_1 - q_2 \overline{\lambda}_0 ) \Delta_q^{-1/2}
\end{equation}
\begin{equation} \label{deltaq}
\Delta_q = (q_1\ q_2) \mathcal{M}_0^{-1} \left(\begin{array}{c}
q_1\\  q_2 \end{array} \right) = e^{\phi_{0}} (q_2 \chi_0 - q_1)^2 +
e^{-\phi_{0}} q_2^2,
\end{equation}
which gives the $SL(2,\ZZ)$ covariant result
\begin{equation}
\alpha_q = \Delta_q^{1/2} Q.
\end{equation}
The general solution describing a $(q_1, q_2)$ string in the
$\lambda_0$ vacuum now follows. It is given by the metric in eq.~(\ref{metric}) and
\begin{equation}
B^{(i)}_{01} =  (\mathcal{M}_0^{-1})_{ij} q_j \Delta_q^{-1/2} A_q^{-1}
\end{equation}
\begin{equation} \label{lambform}
\lambda = {q_1 \chi_0 - q_2 |\lambda_0|^2  +i q_1 e^{-\phi_{0}}A_q^{1/2}\over
q_1 - q_2 \chi_0   + i q_2 e^{-\phi_{0}}A_q^{1/2}},
\end{equation}
which satisfies $\lambda \rightarrow \lambda_0$.\footnote{The original version of this paper
had $\mathcal{M}_0$ instead of $\mathcal{M}_0^{-1}$ in eq.~(\ref{deltaq}).
The key to getting this right is to note that the charges $(q_1,q_2)$ transform
contragrediently to the gauge fields $(B^{(1)}, B^{(2)})$ under $SL(2,\ZZ)$.
I am grateful to S. Roy and R. Gebert for pointing out additional errors in these equations.}

We have now obtained an $SL(2,\ZZ)$ family of type IIB macroscopic strings
carrying $H$ charges $(q_1, q_2)$ and having tensions
\begin{equation}
T_q = \Delta_q^{1/2} T.
\end{equation}
For generic values of $\lambda_0$ one of these tensions is smallest.  However,
for special values of $\lambda_0$ there are degeneracies.  For example, $T_{1,0}
= T_{0,1}$ whenever $|\lambda_0| = 1$.  (More generally, $T_{q_1,q_2}
=T_{q_2,q_1}$ in this case.)
Also, $T_{1,0} = T_{1,1}$ whenever $\chi_0
= - {1\over 2}$.   (More generally, $T_{q_1,q_2}
=T_{q_1,q_1 - q_2}$ in this case.)
Combining these, we find a three-fold degeneracy $T_{1,0} =
T_{0,1} = T_{1,1}$ for the special choice $\lambda_0 = e^{2\pi i/3}$.

Although we have only constructed infinite straight macroscopic strings, there
must be an infinite family of little loopy strings whose spectrum of
excitations can be analyzed in the usual way.  Thus, in ten dimensions the
$(q_1, q_2)$ string should have a perturbative spectrum given by $M^2 = 4\pi
T_q (N_L + N_R)$, where $N_L$ and $N_R$ are made from oscillators in the usual
way.  All the different strings have the same lowest level --- the IIB
supergravity multiplet --- in common.  The excited states are presumably all
distinct, with the excited levels of one string representing states that are
non-perturbative from the viewpoint of any of the other strings.  Of course,
the formula for $M^2$ gives the free-particle spectrum only, which  
is not meaningful for all the strings at the same time, so comparisons 
of massive levels are only
qualitative.  In ten dimensions, the only states in short supersymmetry
multiplets, for which we have good control of the quantum corrections, are those
of the supergravity multiplet itself.  We now turn to the theory compactified
to nine dimensions, because much more of the spectrum is under precise control
in that case.

Consider the $(q_1, q_2)$ IIB string compactified on a circle of radius $R_B$.
Then the resulting perturbative spectrum of this string  has nine-dimensional
masses given by
\begin{equation}
M_B^2 = \left({m\over R_B}\right)^2 + (2\pi R_B n T_q)^2 + 4\pi T_q (N_L +
N_R),
\end{equation}
where $m$ is the Kaluza--Klein excitation number (discrete momentum) and $n$ is
the winding number, as usual.  Level-matching gives the condition
\begin{equation}
N_R - N_L = mn.
\end{equation}
Short multiplets (which saturate a Bogomol'nyi bound) have $N_R = 0$ or $N_L =
0$. (Ones with $N_L=N_R=0$ are `ultrashort'.)
Taking $N_L =0$ gives $M_B^2 = ( 2\pi R_B n T_q +{m/ R_B})^2$
and a rich spectrum controlled by $N_R = mn$.
The masses of these states should be exact, and they should be stable in
the exact theory.  Note that
\begin{equation}
n^2 T_q^2 = [\ell_2^2 + e^{2\phi_{0}} (\ell_2 \chi_0 - \ell_1)^2]e^{-\phi_{0}}
T^2,
\end{equation}
where $\ell_1 = nq_1$ and $\ell_2 = nq_2$.  Any pair of integers $(\ell_1,
\ell_2)$ uniquely determines $n$ and $(q_1, q_2)$ up to an irrelevant sign
ambiguity.  Winding a $(-q_1, -q_2)$ string $-n$ times is the same thing as
winding a $(q_1, q_2)$ string $n$ times.  Thus the pair of integers $(\ell_1,
\ell_2)$ occurs exactly once, with the tension of the string determined by the
corresponding pair $(q_1, q_2)$.

It is known that the IIB theory compactified on a circle is equivalent to the
IIA theory compactified on a circle of reciprocal radius.  Also, it has been
conjectured that the IIA theory corresponds to eleven-dimensional supergravity
compactified on a circle.  The latter may only be true as an
effective theory~ \cite{witten}, 
or conceivably there is a fundamental 11D supermembrane that
gives rise to the IIA string by double dimensional reduction ~\cite{duff}.  In
either case, these facts imply that there should be a correspondence between
the IIB theory compactified on a circle and eleven-dimensional supergravity
compactified on a torus.\footnote{A detailed comparison of the 9D fields
and dualities obtained by compactifying the 11D, IIA, and IIB  
supergravity theories is
given in Ref. ~\cite{bergshoeffa}.}

Let us consider compactification of eleven-dimensional
supergravity on 
a torus with modular parameter $\tau = \tau_1 + i\tau_2$.  The
Kaluza--Klein modes on this torus are described by wave functions
\begin{equation}
\psi_{\ell_{1},\ell_{2}} (x,y) \sim \exp \left\{{i\over R_{11}} \left[{x}
\ell_2 + {1\over \tau_2} y (\ell_1  - \ell_2 \tau_1)\right]\right\} \quad
\ell_1,\ell_2 \in \ZZ.
\end{equation}
Letting $z =  (x + iy){/2\pi R_{11}}$, 
$\psi_{\ell_{1},\ell_{2}}$ is
evidently invariant under $z \rightarrow z + 1$ and $z \rightarrow z + \tau$.
The contribution to the nine-dimensional mass-squared is given by the
eigenvalue of $p_x^2 + p_y^2 = - \partial_x^2 - \partial_y^2$.  Let us try to
take the supermembrane idea ~\cite{bergshoeffb} seriously, and suppose that
it has 
tension (mass/unit area) $T_{11}$.  Wrapping it $m$ times on a torus of
area $A_{11}$ gives a contribution to the mass-squared of $(m A_{11}
T_{11})^2$.\footnote{The integer $m$ 
is the product of two integers defining the map of
a torus onto a torus. Agreement with the IIB result requires that
different maps giving the same $m$ be identified. I am grateful to
A. Strominger for asking me to clarify this point.}
The area of the torus (evaluated in 
the eleven-dimensional metric) is
$A_{11} = (2 \pi R_{11})^2 \tau_2$.  Therefore, states with wrapping number $m$
and Kaluza--Klein excitations $(\ell_1, \ell_2)$ have nine-dimensional
mass-squared (in the eleven-dimensional metric)
\begin{equation}
M_{11}^2 = \Big(m (2 \pi R_{11})^2 \tau_2 T_{11}\Big)^2 
+ {1\over R_{11}^2} \Big(\ell_2^2 +
{1\over \tau_2^2} (\ell_1  - \ell_2 \tau_1)^2 \Big) + \dots~~,
\end{equation}
where the dots represent membrane excitations, which we do not know how to
compute.  This is to be compared to the equations (19) 
and (21) for $M_B^2$, allowing
$M_{11} = \beta M_B$, since they are measured in different metrics.  Agreement
of the formulas is only possible if the vacuum modulus $\lambda_0$ of the IIB
theory is identified with the modular parameter $\tau$ of the torus.  Since
$SL(2, \ZZ)$ is the modular group of the torus, this provides strong evidence
that it should also be the duality group of the IIB string.  In addition, the
identification $M_{11} = \beta M_B$ gives
\begin{equation} \label{RsubB}
R_B^{-2} = T T_{11} A_{11}^{3/2},
\end{equation}
\begin{equation}
\beta^2 = 2\pi R_{11} e^{-\phi_0 /2}  T_{11}/T.
\end{equation}
These identifications imply predictions for the spectrum of membrane
excitations -- at least those that give short supermultiplets.

It is also interesting to explore how the type IIA string fits into the story.
Compactification on a circle of radius $R_A$ gives nine-dimensional masses
\begin{equation}
M_A^2 = \left({\ell_2\over R_A}\right)^2 + 
(2\pi R_A m T_A)^2 + 4\pi T_A (N_L + N_R).
\end{equation}
This can be matched to 
the formula for $M_B^2$ in eq. (19) in the special case $(q_1,
q_2) = (1,0)$, $\chi_0 = 0$. 
Since $M_A$ 
and $M_B$ might be measured in different metrics, we set
$M_B = \gamma M_A$. 
Comparison of the formulas
then gives the $T$-duality relation~\cite{dai}
\begin{equation} \label{RARB}
R_A R_B = (2\pi \gamma T_A)^{-1}
\end{equation}
and $T_A = \gamma^{-2} e^{-\phi_{0}/2} T$.
Using the factor $\beta\gamma$ to convert 
$R_A$ to the eleven-dimensional metric 
and substituting eqs.~(\ref{RsubB}) -- (\ref{RARB}) gives
$R_A^{(11)} = R_A/\beta\gamma = R_{11}$ .
The meaning of this is that when $\chi_0 = \tau_1 = 0$, the torus is a
rectangle (with opposite sides identified) and the sides have lengths $2\pi
R_{11}$ and $2\pi r = 2 \pi R_{11} \tau_2$.  $r$ is the radius of the
circle that
takes eleven-dimensional supergravity to the $D=10$ IIA theory, and $R_{11}$
is the radius of circle taking the $D=10$ IIA theory to $D = 9$.
The T duality relating IIA and IIB superstrings in
nine dimensions is best regarded as a duality between a
torus and a circle. Wrappings of the supermembrane on the
torus correspond to Kaluza--Klein excitations of the circle,
and windings of an $SL(2,\ZZ)$ family of strings on the circle
correspond to Kaluza--Klein excitations of the torus.

The comparison just presented only shows how to interpret some of the
type IIB states from the type IIA viewpoint. Generalizing to
arbitrary $\lambda_0$ is not a problem. More interesting is the issue of
how to describe states with $\ell_2 \neq 0$. Even though we have
found an infinite family of IIB strings, the IIA string
in $D=10$ is unique. As has been noted by
Townsend \cite{bergshoeffb}
and Witten \cite{witten}, $D=10$ IIA states with $\ell_2 \neq 0$
arise as zero-branes (sometimes called `extremal black holes')
and are not present in the perturbative IIA string spectrum. The corresponding
$D=9$ states can be constructed by taking a periodic array of these
zero-branes
in $D=10$ before identifying $x^1$ and $x^1 + 2\pi R_A$. This 
construction is possible,
because the zero-brane solutions 
saturate a Bogomol'nyi bound and therefore satisfy a
no-force condition. A rich spectrum of such zero-branes in $D=10$ is
required in order to match up with the $D=9$ spectrum that we have found
from the IIB analysis.

In Ref. ~\cite{witten}, Witten found that $r \sim e^{2 \phi_{A}/3}$, where
$\phi_A$ is the dilaton of the $D = 10$ IIA theory.  Let us explore how this
result meshes with what we have found.  The first thing to note is that $\phi_A$
is not the same thing as $\phi_0$.  As we have just learned (for a rectangular
torus), $\phi_0$ is related to ratio of the lengths of the two periods of the
torus.  
$\phi_A$, on the other hand, is defined in ten dimensions and cannot know
anything about the radius of a circle defining a subsequent reduction to nine
dimensions.  Witten derived the formula for $r$ by comparing the masses of
IIA states in ten dimensions with non-zero $\ell_2$ charges
that saturate a Bogomol'nyi bound to those of Kaluza--Klein
excitations of 11D supergravity compactified on a circle of radius $r$.  
An alternative
approach is to consider wrapping the 11D supermembrane on a circle of radius
$r$ to give a type IIA string with tension 
$2\pi r T_{11}$ \cite{duffa}.\footnote{It is somewhat puzzling why a
supermembrane can wrap around a two-torus any number of times, but it
can wrap around a circle only once \cite{becker}.}  
This tension is measured in
units of the eleven-dimensional metric $g^{(11)}$.  It can be converted to the
ten-dimensional canonical metric $g_{CAN}^{(10)}$ and the
string metric of the IIA string $g_{ST}^{(10)}$ using
\begin{equation}
g^{(11)} = e^{-\phi_{A}/6} g_{CAN}^{(10)} = e^{-2\phi_{A}/3} g_{ST}^{(10)} .
\end{equation}
Denoting the IIA string tension in the IIA string metric by $T_{ST}$,
we deduce that
\begin{equation}
T_{ST} = 2 \pi r T_{11} e^{-2\phi_{A}/3}.
\end{equation}
In this way, we have confirmed that $r \sim e^{2\phi_{A}/3}$ and even
determined the constant of proportionality,
though it is unclear to me precisely how $T_{ST}$ is related to $T$.
Using the formulas given above,
$\phi_A$ can be related to the parameters of the compactified IIB string.

To conclude, there is an infinite family of type IIB superstrings labeled by a
pair of relatively prime integers, which correspond to their $H$ charges.  Any
one of the strings can be regarded as fundamental with the rest describing
non-perturbative aspects of the theory. This family of strings has tensions
given by the $SL(2,\ZZ)$ covariant expression in eqs. (15) and (18).
Compactifying to $D=9$ and identifying with eleven-dimensional supergravity
compactified on a torus requires equating the modulus $\lambda_0$ of the IIB
theory and the modular parameter $\tau$ of the torus, which is strong evidence
for $SL(2,\ZZ)$ duality.  Other aspects of the spectrum are consistent with a
supermembrane interpretation. 

I would expect that, just as we found for strings, 
five-brane solutions of the D=10 IIB theory (for a specified
vacuum $\lambda_0$) also form an $SL(2,\ZZ)$
family labeled by a pair of relatively prime integers. (See Ref.
\cite{bergshoeffc} for a recent discussion of five-brane solutions.)
The self-dual three-brane \cite{horowitz}, on the other hand, 
should be unique.

I wish to acknowledge helpful discussions with A. Dabholkar, 
J. Gauntlett, G. Gibbons, M.
Green, G. Horowitz, C. Hull, and A. Sen.  I am grateful to the Aspen Center for
Physics, where most of this work was done.

Note added: After completion of this paper I learned that the
interpretation of the $SL(2,\ZZ)$ duality group of type IIB
superstrings as that of a torus has also been pointed out by Aspinwall
\cite{aspinwall}.

\vfill\eject

\end{document}